\documentclass[a4paper,11pt]{article}
\usepackage{pos}

\title{A Kinetic Study of the Saturation of the Bell Instability}
\ShortTitle{The Saturation of the Bell Instability}

\author*[a]{Georgios Zacharegkas}
\author[a,b]{Damiano Caprioli}
\author[c]{Colby Haggerty}
\author[a]{Siddhartha Gupta}
\affiliation[a]{Department of Astronomy and Astrophysics, The University of Chicago, 5640 S Ellis Ave, Chicago, IL 60637, USA}
\affiliation[b]{Enrico Fermi Institute, The University of Chicago, 5640 S Ellis Ave, Chicago, IL 60637, USA}
\affiliation[c]{University of Hawaii, Institute for Astronomy}

\emailAdd{gzacharegkas@uchicago.edu}

\abstract{The nonresonant cosmic ray instability, predicted by Bell (2004), is thought to play an important role in the acceleration and confinement of cosmic rays (CR) close to supernova remnants. Despite its importance, the exact mechanism responsible for the saturation of the instability has not been determined, and there is no first-principle prediction for the amplitude of the saturated magnetic field. 
Using a survey of self-consistent hybrid simulations (with kinetic ions and fluid electrons), we study the non-linear evolution of the Bell instability as a function of the parameters of the CR population.
We find that saturation is achieved when the magnetic pressure in the amplified field is comparable to the initial CR momentum flux.}

\FullConference{%
  *** 37th International Cosmic Ray Conference (ICRC2021), ***\\
  *** 12-23 July 2021 ***\\
  *** Berlin, Germany - Online ***
}

\begin{document}
\maketitle

%%%%%%%%%%%%%%%%%%%%%%%%%%
%% Introduction
%%%%%%%%%%%%%%%%%%%%%%%%%%
\section{Introduction}\label{sec:intro}

High energy cosmic rays (CR), while relatively few by number, make up a significant fraction of the energy budget of the interstellar medium. Collisionless shock waves associated with supernova remnants (SNRs) are believed to be the primary source of galactic CRs (up to "knee" energies of $\sim 10^{15}$ eV), through the diffusive shock acceleration (DSA) mechanism \citep{bell78a}. However, for efficient CR acceleration through DSA, CRs must be confined close to the shock, which requires the presence of very strong, turbulent magnetic fields. Strong magnetic turbulence and CR acceleration are thus closely related and the study of the excitation, growth and saturation of such strong magnetic fields is crucial to explain the origin of high-energy CRs.

\cite{winske+84} and \cite{bell04} found that in systems with a sufficiently strong CR current, modes with wavelengths significantly smaller than the CR gyroradius could be excited; these modes are different from those driven by the resonant cosmic-ray instability, which is caused by CRs in gyroresonance with Alfv\'{e}nic modes \citep{kulsrud+69}. 
This \emph{nonresonant streaming instability}, which we refer to as the "Bell instability", has important implications, e.g., for DSA of Galactic CRs in SNRs \citep{caprioli+14b,bell+13,cardillo+15}. 

In \cite{gargate+10} hybrid (with kinetic ions and fluid electrons) PIC simulations were used to follow the instability well into the non-linear regime where the saturation takes place. They deduced that the saturation occurs due to the deceleration of CRs and simultaneous acceleration of the background plasma, which reduces the net CR current. This deceleration of CRs was also seen in earlier works \cite{lucek+00,riquelme+09}.

In this paper we study the Bell instability using {\it dHybridR} simulations \citep{haggerty+19a}, with kinetic ions and fluid electrons in the plasma, by injecting CRs in the simulation box at a constant rate. 
We mostly focus on 2D simulations, validated by a few 3D runs. 
We study the conditions that lead to the saturation of the instability and derive a formula to predict the level of the final magnetic field as a function of the initial conditions in our simulations.

%%%%%%%%%%%%%%%%%%%%%%%%%%
%% Theory discussion
%%%%%%%%%%%%%%%%%%%%%%%%%%
\section{Bell Instability: Linear Theory}\label{sec:theory}

We consider a population of streaming CRs, with number density $n_{\rm cr}$, that carry an electric current $\mathbf{J}_{\rm cr} = e n_{\rm cr} \mathbf{v}_{\rm cr}$, where $e$ is the proton charge and $\mathbf{v}_{\rm cr}$ is the drift velocity of the injected CR population; 
CRs travel parallel to a background magnetic field $\mathbf{B}_0$ through a plasma of number density $n_g$.
For a sufficiently energetic CR beam \citep{bell04,amato+09}, the Bell instability has a maximum growth rate of
\begin{equation}\label{eqth:gamma_max Bell}
	\gamma_{\max} = k_{\max} v_{\rm A,0} = \frac{1}{2} \left( \frac{n_{\rm cr}}{n_g} \right) \left( \frac{v_{\rm cr}}{v_{\rm A,0}} \right)\Omega_{ci} \; ,
\end{equation}
at the fastest growing mode with wavelength $\lambda_{\max} = 2\pi / k_{\max}$, where
\begin{equation}\label{eqth:k_max Bell}
	k_{\max} = \frac{4\pi}{c} \frac{J_{\rm cr}}{B_0} = \frac{1}{2} \left( \frac{n_{\rm cr}}{n_g} \right) \left( \frac{v_{\rm cr}}{v_{\rm A,0}} \right) d_i^{-1} \;.
\end{equation}
Here $v_{\rm A,0} = B_0/\sqrt{4\pi m n_g}$ is the Alfv\'{e}n speed based on the background plasma density and magnetic field, $\Omega_{ci}=eB_0/(mc)$ is the ion gyrofrequency, $m$ is the ion mass, $c$ is the speed of light, and $d_i = v_{\rm A,0}/\Omega_{ci}$ is the ion skin length.

\begin{table}[t]
	\centering
	\begin{tabular}{c c c c c c}
		\hline
		{Run} &
		{$n_{\rm cr}$} &
		{$p_{\rm cr}$} &
		{$p_{\rm iso}$} &
		{$L_x \times L_y (\times L_z)$} &
		{$N_d$} \\
		{} & 
		{$(n_g)$} &
		{$(mv_{\rm A,0})$} &
		{$(mv_{\rm A,0})$} &
		{($d_i^2$ or $d_i^3$)} &
		{} \\
		\hline \hline
		$\mathcal{B}$ & $10^{-3}$ & $10^3$ & $1$ & $10^3 \times 10^3$ & $2$ \\
		$\mathcal{C}1$ & $10^{-3}$ & $1.5 \times 10^{3}$ & $1$ & $10^3 \times 10^3$ & $2$ \\
		$\mathcal{C}2$ & $10^{-3}$ & $5 \times 10^{2}$ & $1$ & $10^3 \times 10^3$ & $2$ \\
		$\mathcal{C}3$ & $10^{-3}$ & $2 \times 10^{3}$ & $1$ & $10^3 \times 10^3$ & $2$ \\
		$\mathcal{C}4$ & $10^{-3}$ & $7.5 \times 10^{2}$ & $1$ & $10^3 \times 10^3$ & $2$ \\
		$\mathcal{C}5$ & $10^{-3}$ & $10^{4}$ & $1$ & $10^3 \times 10^3$ & $2$ \\
		$\mathcal{C}6$ & $10^{-3}$ & $5 \times 10^{3}$ & $1$ & $10^3 \times 10^3$ & $2$ \\
		$\mathcal{C}7$ & $10^{-3}$ & $3 \times 10^{3}$ & $1$ & $10^3 \times 10^3$ & $2$ \\
		$\mathcal{C}8$ & $10^{-3}$ & $8 \times 10^{3}$ & $1$ & $10^3 \times 10^3$ & $2$ \\
		$\mathcal{C}9$ & $10^{-3}$ & $2 \times 10^{2}$ & $1$ & $10^3 \times 10^3$ & $2$ \\
		$\mathcal{C}10$ & $10^{-3}$ & $3 \times 10^{2}$ & $1$ & $10^3 \times 10^3$ & $2$ \\
		$\mathcal{C}11$ & $10^{-3}$ & $4 \times 10^{2}$ & $1$ & $10^3 \times 10^3$ & $2$ \\
		$\mathcal{H}1$ & $10^{-3}$ & $10^{3}$ & $10^{3}$ & $10^3 \times 10^3$ & $2$ \\
		$\mathcal{H}2$ & $10^{-3}$ & $10^{3}$ & $5 \times 10^{2}$ & $10^3 \times 10^3$ & $2$ \\
		$\mathcal{H}3$ & $10^{-3}$ & $10^{3}$ & $10$ & $10^3 \times 10^3$ & $2$ \\
		$\mathcal{H}4$ & $10^{-3}$ & $10^{3}$ & $50$ & $10^3 \times 10^3$ & $2$ \\
		$\mathcal{H}5$ & $10^{-3}$ & $10^{3}$ & $10^2$ & $10^3 \times 10^3$ & $2$ \\
		$\mathcal{H}6$ & $10^{-2}$ & $10^{2}$ & $10^3$ & $10^3 \times 10^3$ & $2$ \\
		$\mathcal{D}1$ & $10^{-3}$ & $10^{3}$ & $1$ & $10^3 \times 50$ & $1$ \\
		$\mathcal{D}3$ & $10^{-3}$ & $10^{3}$ & $1$ & $10^3 \times 200 \times 200$ & $3$ \\
		\hline
	\end{tabular}
	\caption{List of the simulations used in this work. The second column shows the CR number density, $n_{\rm cr}$, normalized by the gas number density $n_g$; column $p_{\rm cr} = \gamma_{\rm cr} m v_{\rm cr}$ corresponds to the parallel CR momentum component; the initial isotropic momentum $p_{\rm iso} = \gamma_{\rm iso} m v_{\rm iso}$ of the CR population is shown in the fourth column; the last two columns report the size of the simulation box in skin depth, $d_i$, and the dimensionality of the simulation, respectively. 
	The speed of light is always fixed to $c=100 \; v_{\rm A,0}$. 
	The first row (simulation $\mathcal{B}$) is our benchmark; cold beam cases with $p_{\rm cr} \gg p_{\rm iso}$ are denoted by "$\mathcal{C}$", while hot cases with $p_{\rm iso} > p_{\rm cr}$ are denoted by the letter $\mathcal{H}$. 
	The two last rows correspond to control 1D and 3D runs.}
	\label{tab:sims}
\end{table}

%%%%%%%%%%%%%%%%%%%%%%%%%%
%% Hybrid simulations
%%%%%%%%%%%%%%%%%%%%%%%%%%
\section{Hybrid Simulations: Setup and Time Evolution}\label{sec:simulations}

To study the saturation of the Bell instability we preform kinetic hybrid simulations with {\it dHybridR}, in which the ions are treated as macro-particles governed by the relativistic Lorentz force law and the electrons are a massless, charge neutralizing fluid \citep{haggerty+19a}. 
In our simulations the background magnetic field is oriented along the $x$-direction. Henceforth, we will refer to this as the {\it parallel} direction denoted by the symbol "$||$", while the directions perpendicular to it, i.e., the $y\ \&\ z$ directions, will be the {\it perpendicular} denoted by "$\perp$".

We have run a large number of 2D simulations to study how the different CR parameters affect the saturation. In each system CRs were injected either as a {\it cold beam} or as a {\it hot drifting} population of particles into the simulation box from the left boundary. The CR distribution is defined by a drift velocity in the $x$ direction, of magnitude $v_{\rm cr}=p_{\rm cr}/(m\gamma_{\rm cr})$, plus an isotropic momentum of modulus $p_{\rm iso}=m\gamma_{\rm iso}v_{\rm iso}$. The Lorentz factors $\gamma_{\rm cr}$ and $\gamma_{\rm iso}$ are respectively defined based on $p_{\rm cr}$ and $p_{\rm iso}$. The full list of parameters of the simulations we considered in this work is shown in Table \ref{tab:sims}. The CR ($s={\rm cr}$) and plasma ($s={\rm gas}$) species have pressure given by the tensor
\begin{align}\label{eqsim:pressure tensor}
	P_{s,ij}(t) = \int d^3 \mathbf{p}_s f_s(\mathbf{p}_s,t) \; v_{s,i} p_{s,j} \;  ,
\end{align}
where $f_s(\mathbf{p}_s,t)$ is the CR or plasma distribution function in momentum space.

\begin{figure*}[t]
\centering
\includegraphics[width=6.in]{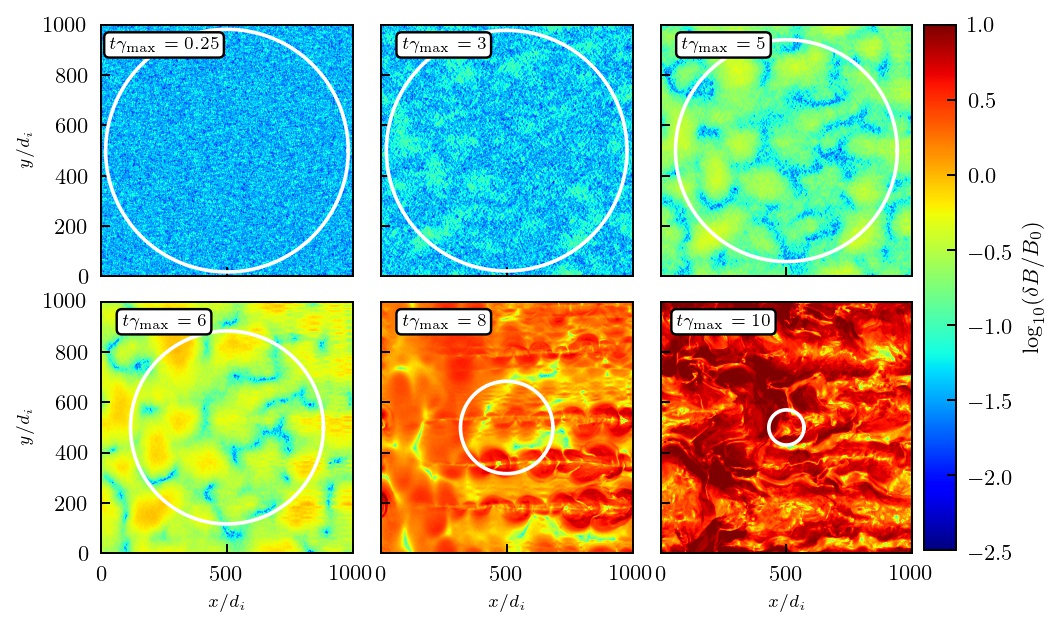}	
\caption{Time evolution of $\delta B = (B_y^2 + B_z^2)^{1/2}$ in our benchmark simulation $\mathcal{B}$ from Table~\ref{tab:sims}. Time has been normalized by $\gamma_{\max}^{-1}$, and the axes represent the physical size of the simulation box in units of skin depth. The color scale indicates the strength of the field in units of $B_0$. The CR Larmor radius, defined in the amplified magnetic field, is represented by the white circles.}
\label{fig:B-field evolution in sim box}
\end{figure*}

Regardless of the physical dimensions of the box,  simulations retain all the components of the velocity and fields. 
Periodic boundary conditions are used for fields and thermal particles, while for CRs the box is periodic in the transverse direction and open the $\pm x$ directions. 
This configuration mimics an astrophysical environment in which energetic CRs are being continuously injected, e.g., by a shock (driven instability). 
We have checked convergence of the field at saturation with the box size, the number of macroparticles per cell, and the grid resolution;
in particular, we found that the final $\delta B$ converges when the parallel and perpendicular lengths are at least one CR gyroradius, defined based on the initial magnetic field.

The various runs in Table~\ref{tab:sims} have been labeled based on whether the injected CR population has a distribution in momentum space similar to that of a cold beam ($\mathcal{C}1$-$\mathcal{C}11$), with $p_{\rm cr} \gg p_{\rm iso}$, or that of a hot drifting population of particles ($\mathcal{H}1$-$\mathcal{H}6$), where $p_{\rm iso} > p_{\rm cr}$. 
Our benchmark 2D run, called $\mathcal{B}$ in Table~\ref{tab:sims}, is also compared with 1D ($\mathcal{D}1$) and 3D ($\mathcal{D}3$) simulations to assess that 2D runs can capture the full non-linear physics of the problem.

The growth of the instability proceeds in agreement with the linear theory \citep{bell04,amato+09}. 
We quickly see structures of amplified magnetic fields forming inside the simulation box, which grow in size, as seen in Figure~\ref{fig:B-field evolution in sim box}. 
Initially, while in the linear phase, these structures grow exponentially over time, but eventually the growth slows down and the field saturates.

\begin{figure}[t]
\centering
\includegraphics[width=\columnwidth]{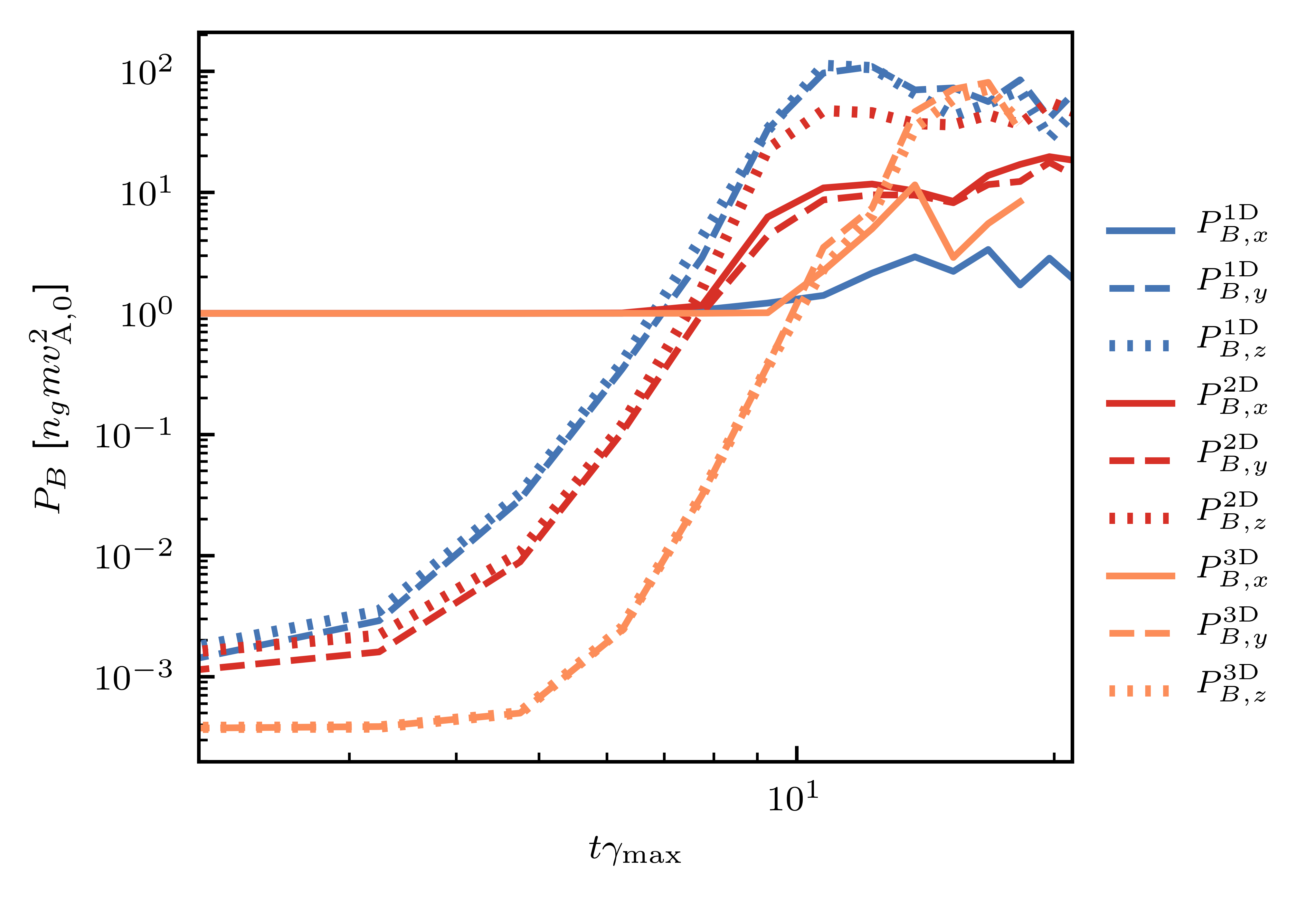}
\caption{
Time evolution of the parallel (solid) and perpendicular (dashed for the $y-$ and dotted for the $z-$direction) components of the magnetic field for our benchmark case, from Table~\ref{tab:sims}, in 1D ($\mathcal{D}1$), 2D ($\mathcal{B}$) and 3D ($\mathcal{D}3$) configurations.
Note how the out-of-plane component $B_z$ in 2D runs matches is representative of both transverse components in 3D runs.
}
\label{fig:1D vs 2D vs 3D}
\end{figure}

The moment where $\delta B / B_0$ exceeds unity marks the starting of CR scattering by the self-generated magnetic field. 
This results in transfer of momentum from the CRs to the background gas: CRs decelerate and the thermal gas is set in motion in the direction of the CR drift. 
As a result, the CR current is reduced in the frame of the gas, which slows the growth of the instability down.
The panels is Figure~\ref{fig:B-field evolution in sim box} also include a white circle that depicts the instantaneous CR Larmor radius, $r_L$, which reduces in size as the magnetic field is being amplified. 
CR scattering becomes significant when $r_L$ becomes comparable to the size of the large magnetic-field structures that form in the simulation box.

\begin{figure}
	\centering
\includegraphics[width=\columnwidth]{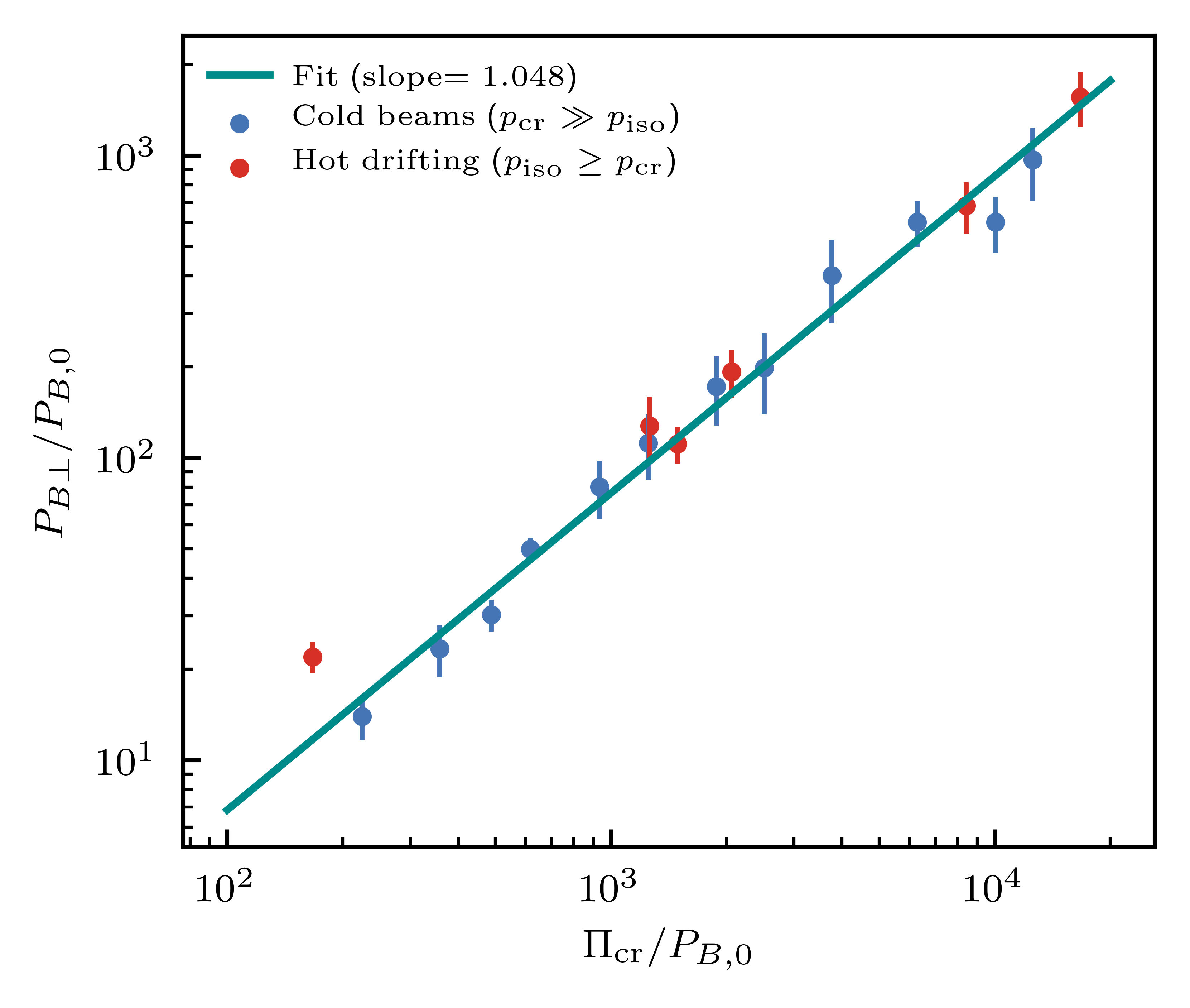}
	\caption{\label{fig:saturation_plot} 
	Saturated magnetic pressure as a function of the anisotropic CR momentum flux density in the CR drift direction. The error bars show the fluctuations on the saturated average value of the magnetic field. The slope of the best-fit line in this log-log plot is close to unity, showing that the two quantities are linearly proportional. Blue (red) points correspond to cold beam (hot drifting) injected CR populations.}
	\label{fig:saturation plot}
\end{figure}

%%%%%%%%%%%%%%%%%%%%%%%%%%
%% Results
%%%%%%%%%%%%%%%%%%%%%%%%%%
\section{Saturation of the Magnetic Field Amplification}\label{sec:Results}

Quantifying the value of magnetic field at saturation in 2D simulations is not a trivial task, since they typically show an anisotropy in the $B$-field in the transverse components, $B_y$ (in plane) and $B_z$ (out of plane). 
Specifically, we find that $B_z$ can exceed in magnitude $B_y$ by a significant amount, depending on the simulation parameters, and it is always the case that $B_z \gtrsim B_y$. 
This makes it hard to assess which component of the magnetic field must be considered when we calculate the saturated value of the field. 
In 1D boxes, i.e. when $L_y \ll L_x$, we find that $B_y \sim B_z > B_x$, and that $B_x \sim B_0$ as expected. When a 3D simulation is considered, we find that $B_y \sim B_z$, similarly to the 1D case, and that the two components are comparable to the $B_z$ one in the 2D box. 
All the above are demonstrated in Figure~\ref{fig:1D vs 2D vs 3D} where we compare simulations $\mathcal{D}1$, $\mathcal{D}3$ and $\mathcal{B}$ (see Table \ref{tab:sims}).
Our test indicates that the $B_z$ component in our 2D simulations is a good proxy for both components of the amplified magnetic field in realistic 3D setups, so in the following we use such an out-of-plane component for quantifying the fiducial value of the saturated magnetic fields in 2D runs. 

Figure \ref{fig:saturation_plot} shows the final, amplified magnetic field pressure, defined based on $B_z$, as a function of the initial anisotropic CR momentum flux density, $\Pi_{\rm cr}$; 
this quantity corresponds to the $01$-component of the stress-energy tensor, $\Pi_{\rm cr} \equiv T^{01}$. 
The plot clearly shows that  $P_{B\perp} / P_{B,0} = \delta B^2/B_0^2 \approx 2 B_z^2 / B_0^2$ and $\Pi_{\rm cr}$ are almost perfectly proportional to each other, which indicates that $\Pi_{\rm cr}$ is a good indicator for the saturated value of the magnetic field.
Rewriting $\Pi_{\rm cr}$ as a function of the initial CR parameters yields the relation
\begin{equation}\label{eq:Pi_CR}
    \frac{\delta B^2}{4\pi} \approx \frac{B_z^2}{2\pi} \approx \Pi_{\rm cr} = n_{\rm cr} \gamma_{\rm iso} \left[ 1 + \left( \frac{p_{\rm iso}}{\gamma_{\rm iso} m c} \right)^{2/3} \right] \frac{p_{\rm cr}^2}{\gamma_{\rm cr}} \; .
\end{equation}
This result corresponds to the long-sought prediction for the final level of magnetic field amplification that an arbitrary anisotropic CR distribution can produce. 
It is possible to show that Eq.~\eqref{eq:Pi_CR} reduces to the ansatz put forward by Bell for a shock that accelerates CRs (Eq.~(28) of \citep{bell04}) and also to the ansatz suggested by Blasi et al.~(Eq.~(6) of \citep{blasi+15}).
Despite some of the features of the non-linear stages of the Bell instability have already been observed in prior kinetic simulations \cite{ohira+09, riquelme+09, niemiec+08, gargate+10}, this is the first time that the heuristic argument about equipartition between the magnetic pressure at saturation and the initial anisotropic CR pressure has been validated by ab-initio kinetic simulations. 

We will present our full analysis and detailed explanation behind this in a forthcoming publication.

%%%%%%%%%%%%%%%%%%%%%%%%%%
%% Bibliography
%%%%%%%%%%%%%%%%%%%%%%%%%%
\bibliographystyle{JHEP}
\bibliography{Total}

\end{document}